# Mobile Phone Forensics: An Investigative Framework based on User Impulsivity and Secure Collaboration Errors

Milda Petraityte[1], Ali Dehghantanha[2], Gregory Epiphaniou[3]
1,2: School of Computing, Science and Engineering, University of Salford- United Kingdom
3: Department of Computer Science and Technology, University of Bedfordshire – United Kingdom
milda.petraityte@hotmail.com, a.dehghantanha@salford.ac.uk,
gregory.epiphaniou@beds.ac.uk

**Abstract**

This paper uses a scenario-based role-play experiment based on the usage of QR codes to detect how mobile users respond to social engineering attacks conducted via mobile devices. The results of this experiment outline a guided mobile phone forensics investigation method which could facilitate the work of digital forensics investigators while analysing the data from mobile devices. The behavioural response of users could be impacted by several aspects, such as impulsivity, smartphone usage and security or simply awareness that QR codes could contain malware. The findings indicate that the impulsivity of users is one of the key areas that determine the common mistakes of mobile device users. As a result, an investigative framework for mobile phone forensics is proposed based on the impulsivity and common mistakes of mobile device users. As a result, an investigative framework for mobile phone forensics is proposed based on the impulsivity and common mistakes of mobile device users. It could help the forensics investigators by potentially shortening the time spent on investigation of possible breach scenarios.

# 1. Introduction

The issues with mobile device security have been under the radar of scholars and security researchers for a while now. As convenient as mobile devices have become, their versatile interaction with the environment and their users have vulnerabilities that have received attention from multiple researches [1]–[7]. For a long time social engineering and flaws with user behaviour online was associated with computers [8]. However, as technology evolves, users now face another challenge, as the mobile devices have evolved to become a powerful tool that is connected to the internet and also cheap enough to be available to a large population of users that may be unable to afford a laptop and broadband service. It is thought that '10% of Americans own a smartphone but do not have broadband at home, and 15% own a smartphone but say that they have a limited number of options for going online other than their cell phone, where 'smartphone dependent' are likely to be people with low income, young adults or less educated ones [9]. Furthermore, Kaspersky Lab reported an intensive growth in mobile malware since 2011 and in 2013 there was more than 148,327 items of mobile malware [3], [10]. Therefore, it is very likely that users will regularly face threats related to mobile devices, yet they are likely to be unable to recognise them because of poor security awareness. Under such circumstances, it is not likely that users will manage the security of their mobile device very well. Another factor that contributes to poor hygiene when it comes to mobile security, is that users do not seem to significantly value their privacy prior to its loss. The proliferation of mobile malware has also increased the attack surface and complexity in this paradigm causing traditional defence mechanisms to adapt slower than the overall attack evolvement. Also, network convergence allowed criminals from different areas and backgrounds to gather into the same space with a variety of attack techniques and motives. This allowed for more de-centralised, multi-staged cyber-attacks that demand a different approach in terms of prediction, detection and response.

Moreover, digital forensics investigators often face challenges investigating a variety of mobile devices that differ from one another and they also have to make sure that the key principles of collecting evidence is adhered to [11]. As a result, researchers come up with a variety of methods and tools that help to solve existing issues, such as suggesting possible investigative frameworks for analysis of devices [6], [12]–[14], applications [2], [15], [16] as well as malware [3], [11], [17]–[19]. However, understanding the factors that influence the behaviour of mobile device users and the common mistakes behind the compromise of their mobile devices can be helpful for the forensics investigators and could potentially shorten the time spent on investigation of possible breach scenarios as forensic and allows forensics analysts to conduct risk-aware investigation. This is especially worthwhile since the digital forensics related to mobile devices is of a growing importance to researchers and scientists [20].

This paper aims to address this issue by conducting a scenario-based role-play experiment that investigates how mobile users respond to social engineering via mobile device and to use the result of this experiment for a guided mobile phone forensics investigation method which could facilitate the work of digital forensics investigators while analysing the data from mobile devices. This experiment also helps to identify the trends and emerging challenges that have an impact on mobile device users and gives suggestions for ways of tackling this problem.

For the purposes of this experiment QR Code images were used to mimic social engineering and phishing fraud as QR codes are potentially perfect tool for such attacks and their malicious use has been already observed in the past [21], [22]. Essentially, QR code could be as dangerous as clicking on a link or downloading an attachment in a phishing email, as the scanned code re-directs the user to a fake website, which could also be contaminated with malware.

The rest of this paper is structured as follows. The first section presents the overview of literature and summarises the previous research that has been conducted and explains how this research positions itself to address some of the gaps in the literature. This is followed by an explanation of the research method and details about the survey questionnaire in the second section. Finally, the last section presents and discusses the results and suggests the investigative framework for digital forensics. Limitations of this research as well as conclusions to what has been presented are outlined in sections six and seven respectively.

2. Review of Related Work

Users trust their mobile devices by using them for a wide variety of purposes with the help of the available applications – from communicating with family and friends to shopping and maintaining their finances online [23], [24]. This makes mobile devices a very good yet poorly protected source of valuable information [25]. Naturally, the more users trust their mobile devices, the more attractive these devices become to attackers with malicious intents [19], [21], [26]. Just like malware for the traditional computers, the malware of mobile devices can be developed to perform a variety of tasks [18], such as collecting valuable information, monitoring user activity or performing tasks on behalf of the user with the help of spyware, worms, rootkits, bots and Trojans [17], [18], [21]. This shifting landscape of the cyberspace is clearly manifested by the number of new users or "electronic citizens" joining the Internet on a daily basis. End users are now being actively targeted using the emergence of new mobile devices and applications as a core attack vector. Technical solutions have proven only partially effective with evolving threats rendering static security controls ineffective as more users use their own IT devices.

In spite of efforts on protecting users against social engineering and phishing emails in traditional computing platforms [8], [27], there is very little attention paid for security awareness of mobile device users. Many users jailbreak their devices with no apparent reason of why this should not be done, as it looks like a third party provides a legitimate software, only outside the app store and it may look like there is nothing wrong with that [28], [29]. In fact, users may even think that the app stores actually put unnecessary constrains on downloading and using the applications, while somewhere else they exist for free and with added functionalities [29]. Users, however, are unaware that malware is what usually comes as an added feature too with significant privacy and financial losses [17], [28], [30]. Users may feel that access lock on the device is an inconvenience rather than protection [31] as the common understanding about security is that there is nothing to hide [32], yet it shows that users are unaware of what happens when the device is lost or stolen without having any access protection on it. Due to software development malpractices software developers also create applications that request too many access permissions to the information on the mobile device which could lead to exploitation of the device [4], [18]. However, users hardly ever pay attention to what access is required by the applications as access provisioning is like terms and conditions of software usage that nobody cares to read [29]. Moreover, due to limited screen size of mobile devices, it may be difficult to see all the content that pops up as potential fraud, only a part of a web address may be visible which eventually make it difficult to differentiate between legitimate and malicious contents [8], [29].

QR codes are originally created for a legitimate reason and has a useful purpose, however attackers have realised that QR codes could serve as a perfect social engineering tool which can be used for redirection to fake websites and installation of malware onto user's mobile devices [21], [22]. While QR codes are fairly new feature in the world of technology and their popularity is growing steadily [21], [33], they are bringing up a new route for malware to spread. QR codes are often associated with promotions, discounts or coupons that users expect to receive when they interact with the QR code and the destination website [33]. This is an ideal way to spread fraud as users actually expect to receive something by scanning a QR code, therefore the request to install a malicious application could be perceived as part of a reward process [22], [33]. As a consequence, mobile device users could be particularly susceptible to frauds utilising QR codes.

On the other side there is a lack of unified approach for the investigation of mobile devices and hardly any consideration of user behaviour that led to the compromise of a mobile device [6]. Many researchers focused on collecting evidences of forensics value from mobile phone based on potential location of evidences. Other forensics investigators, especially those investigating iOS platform suggested to collect and preserve data from cloud platforms connected to the mobile device [2], [14], [16], [34]–[36]. However, there is yet any approach to conduct investigation of mobile devices based on common security risks that mobile device users are facing because of their own mistakes.

This paper analyses the role that the users' behavioural factors play in the overall users' perception of mobile device security and then propose a guideline for conducting forensics investigation of mobile devices on the basis of common security mistakes of mobile users.

### 3. Experiment Design

An online web-based questionnaire was created to assess the user interaction with QR codes. It was accessible through an online invitation from the researcher. The questionnaire was divided into three parts. The first part was assessment of 10 QR code images, which were carefully created by the research team. Half of these images contained a genuine URL and the other half was bogus URLs. The 10 QR codes were carefully selected to represent a range of topics that are associated with common day to day activities, such as shopping, advertisement, product labelling and sites for information. The second part of the questionnaire contained demographic information about the participants in order to evaluate their familiarity with smartphone security. It consisted of age, gender, current employment status and level of education. The third part of the questionnaire was used for cognitive reflection test (CRT) to measure the cognitive impulsivity of the test participants. The reason for arranging these parts in such way is to ensure that the demographic questions would not alert the participants that they were involved in a mobile security study until after they had completed their assessment of the 10 QR codes.

A total of 100 participants consisting of university students, lecturers and staff participated in the experiment. The participants were divided into two groups to complete questionnaire in two different sessions. Each session took around 30 minutes and were facilitated by a member of the research team who explained the purpose of the study to participants. In the first session participants were informed that they were participating in a study that aims to understand the user behaviour with regards to QR code without explaining about its possible malicious use. In the second session participants were informed that the study investigates potentials for exploiting some aspects of user's security using QR codes and what is the definition of an 'exploit'. The users were asked to act on behalf of a made up character and their task was to determine if the QR code was genuine or not and, if they could, provide the reasons why the code was fake in their opinion. The reason of this approach was to measure the factor of users' expectancy and how does that affect the results, if at all.

The research aimed to analyse the following variables:

1) **Familiarity with the security of smartphones.** The participants were not directly asked to rate their familiarity with mobile devices, as their judgement would only show their own opinion. The variable was created using the Predictive Analytics Software (PASW) syntax coding to predict the average score of the demographic questionnaire.

2) **Cognitive impulsivity.** To measure that participants completed the CRT which consisted of three questions. Due to the nature of CRT, the participants had to think before providing an answer as the most obvious response was not the correct one. Participants received 1 score for a correct answer while the higher was the score, the better was the person in controlling the impulsivity. This test was chosen as it is likely that the predictive validity of this measure is equal or above other cognitive tests that claim to measure cognitive impulsivity. This measure was included into this study to investigate if impulsivity could lead to wrong identification of fake or genuine QR code.

3) **Behavioural response.** Participants were given the question with answers below. The way they responded was used to measure their behavioural response.

How would you deal with this QR code?

a) Scan the QR code and click on the URL
b) Scan the QR code and verify the URL
c) Scan the QR code and leave it
d) Never scan the QR code

Two behavioural answer variables of each participant were calculated using the previously used PASW software, both for a genuine QR code behaviour score and a genuine QR, as follows.

If the QR code being evaluated as genuine:
- The participant would score 10 for response;
- 1 5 for response;
- 1 2.5 for response; and
- 1 0 for response.

If the QR code being evaluated as fake:
- The participant would score 0 for responses;
- 1 2.5 for response;
- 1 5 for response; and
- 1 10 for response.

Participants could score a maximum of 100 and a minimum of 0 for each type of QR code.

### 4. Results and discussion

Table 2 sums up the average scores for both categories of managing fake QR code and the genuine one. The results show that those participants that knew about the purpose of the questionnaire and the study scored higher than those who did not know what the questionnaire was for. Based on these findings, there was a small difference between the two groups of participants. Overall, genuine QR code was managed better than the fake one.

| Description | Mean Score of Questionnaire Findings |
|---|---|
| Total Point Accumulated for 100 Participants | 4611 |
| Total Score for 100 Participants | 329.326 |
| Mean Score for 1 Participant | 3.29 |

**Table 2: Result of Questionnaires**

## 4.1 Familiarity with smartphones security

According to the findings of the questionnaire, the more familiar users were with smartphone security, the better they dealt with the QR code assessments. This finding was statistically significant and somewhat predictable as people who are familiar with smartphone security are likely to be more aware of the risks and consequences associated with fake QR code rather than people who do not use such devices very often.

|  | **Not Informed** | **Informed** | **Total** |
|---|---|---|---|
| **Fake QR Code** | 22.1 | 29.2 | 25.65 |
| **Genuine QR code** | 30.65 | 31.1 | 30.875 |

Table 3: Average behavioural score for managing QR codes

Table 2 shows that the total score of the online questionnaire survey is 3.29 out of 5. It is fair to conclude that the majority of the participants were familiar with smartphone security. This had no significant impact for dealing with the genuine QR codes but as it is shown in Table 3 it impacted the way the participants treated the fake QR codes.

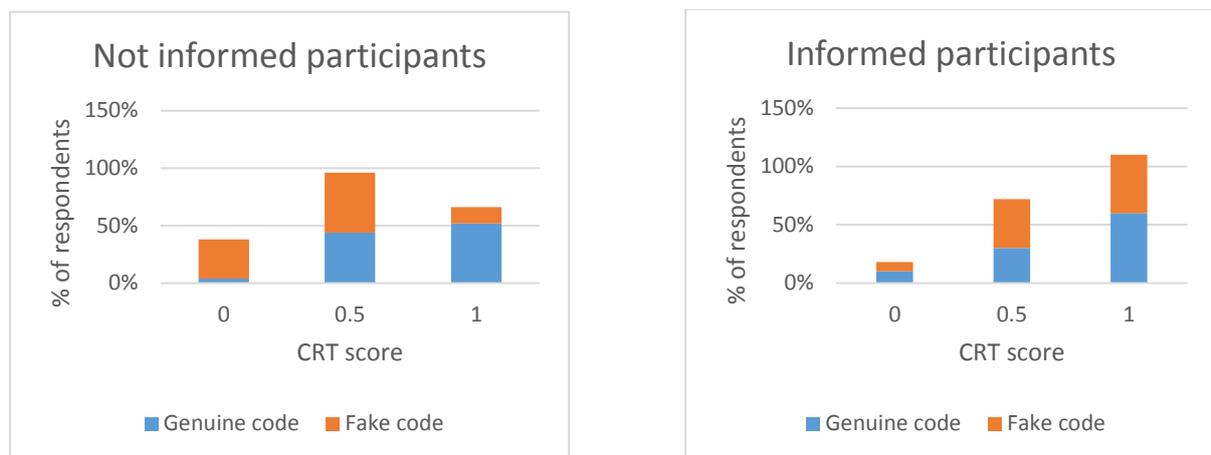

Figure 1: The comparison of CRT scores between the participants that were informed and not informed about the survey and the impact of CRT on how the participants managed the codes.

## 1.1 Cognitive impulsivity

Less impulsive people who did not know what the purpose of this test was (those with a higher CRT score) responded better, as shown in the diagrams below (Figure 1). Participants with higher CRT scores were less likely to click on the URL held inside the fake QR code. The cognitive impulsivity did not show any significant difference on the participants who were informed what the study was about.

## 5. Forensics Investigation Guideline

The performed experiment shows how susceptible users are to downloading and installing a piece of software on their mobile device. Even those participants who knew that they should be careful about the QR codes that were presented to them still did not pay enough attention while making decisions in using the mobile device. This suggests that apart from user education in smartphone security, cognitive impulsivity is the key component in the decision making of users. A logical assumption suggests that the highest potential to act without a second thought arises when a user experiences any kind of social engineering or when they do not pay enough attention while performing other tasks. Depending on the method of the attack a user may be exposed to some social engineering attack attempts before they get tricked into downloading and installing a malware. It could also be that a user does not have a full view of the advertisement or a pop-ups that open together with a legitimate website and could accidentally click on it while trying to switch it off. However, a lot of the common mobile device user mistakes come from conscious decision making and owner's actions.

The suggested forensics investigation guideline that is shown in Figure 2 prioritises forensics investigator actions in regard to common the mistakes that arise from social engineering of end-users, where gradient colours indicate the level of social engineering involved and the next step in the flowchart.

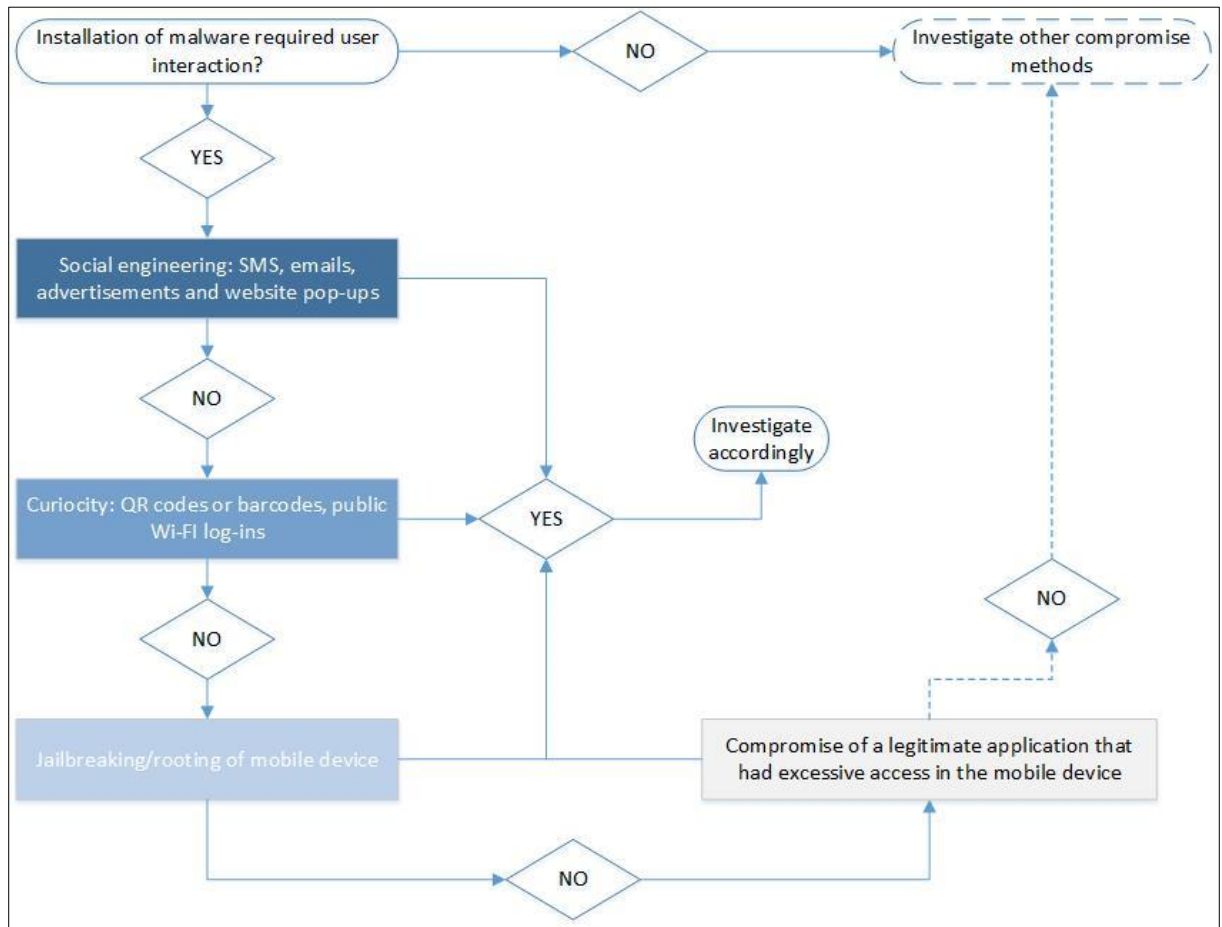

**Figure 2: Suggested forensics investigation guideline based on users' common security mistakes**

Social engineering is understood as tricking someone into 'breaking security policies', where spam is an example of social engineering [37]. Lately there are various types of spam, such as junk emails, Phishing, SMShing, annoying advertisements and pop-ups while browsing the Internet [38]. All types of this spam is relevant to mobile devices and their users when mobile devices allow website content. A user may accidentally click on any of them simply because of their design and lack of full visibility on the screen of a mobile device. Also, it is possible to accidentally click on a pop-up or advertisement just by trying to switch it off and get rid of it as well as to be tricked into thinking that it is a part of a legitimate website that a user tried to open. It is probably the most popular way for malware to spread and attack mobile device users. For this reason as well as potentially the highest likelihood of cognitive impulsivity impact it is suggested to consider this group of items first when conducting a digital investigation.

Second comes a group of actions that contain social engineering, yet a user must make a conscious decision to perform an action. Those are likely to be mistakes that user perform out of curiosity, convenience or benefit. Logging into a compromised public Wi-Fi, a scan of a QR code or a barcode could be a few examples of such behaviour that may lead to the exploitation of the mobile device.

The next step is to check if a user has jailbroken the mobile device. Users jailbreak their mobile devices seeking a variety of benefits without realising that the third party software may contain malware. This action requires a conscious decision and is therefore one of the least impulsive actions that a user may take. Information sharing about third party software repositories and de-centralised threat management systems is more prominent than ever, which contrast to the traditional approaches in cyber/computer security principles related to restrictions in input data and controlling human behaviour. The paradigm is shifted towards monitoring and studying human aspects in users' interaction with their mobiles as part of the formal actions and decision making processes within adversarial clusters,

The last step that is to detect if a user is using legitimate applications that require excessive permissions on a mobile device and the mobile device gets exploited due to compromise of such application, for example via infected application update. Users hardly understand the permissions which are required by various applications and what they are given access to. Possibly even less of users think twice before installing such applications or at least consider disabling the unnecessary access requests.

These common mistakes of mobile device users contain a certain level of impulsivity which, according to the role-play experiment is an important factor that leads to the highest likelihood of mobile device exploitation and end user security flaws. Certainly, other ways of exploiting mobile devices that do not require user interaction are also possible, yet they should be analysed using other available investigation methods.

6. **Limitations**

Similar to any other studies, this study has several limitations as well. Firstly, QR codes is only one threat among others in the spectrum of mobile device exploit possibilities and there could be other equally valid threats associated with mobile devices. However, the phishing-type of social engineering that users are used to seeing as emails can be well replicated by QR codes on mobile devices and therefore QR codes served well for the purpose of the experiment.

Secondly, the study was based on a role play experiment that required users to analyse the given information and provide an advice. However, participants were not required to scan QR code from an unknown source, website or station and therefore results may not imply real-world scenarios.

Moreover, the participants of this study did not have the contextual information that would often influence the decision making. For example, they did not know what was the occupation of the fictitious character on behalf of which they were acting, neither how technology-savvy that person is, nor whether she liked to use the QR code technology.

Another limitation of this research relates to the sample of participants. The total of 100 participants in this study were all members of one university – students, lecturers and staff. A sample of this kind is likely to be biased in terms of age, education, familiarity with mobile security apps and various other factors. Therefore, the conclusions of this experiment are not necessarily representative of the general population of the mobile device users.

## 7. Conclusion and Further Research

The aim of this paper is to conduct a scenario-based role-play experiment that investigates how mobile users respond to social engineering via mobile device and to use the result of this experiment for a guided mobile phone forensics investigation which could facilitate the work of digital forensics investigators while analysing the data from mobile devices. For the purposes of this research the fraud was a QR code and the findings indicate that the genuine QR codes were managed better than the fake ones and this did not depend on how much informed the participants were about the experiment.

Although informed participants performed better in managing both types of QR codes, they were significantly better in managing the fake ones. This implies that educated mobile device users are much better in managing security of their phones. The users should be continuously reminded that the QR codes could contain malware that may infect their devices, yet it is only one aspect of mobile device security. This experiment simply shows that increased user awareness helps to improve secure user behaviour.

This research also analysed few other factors that have impact on users when they face a QR code. The survey results indicate that controlled impulsiveness is associated with improved performance in managing fake QR codes in users who were not informed about the purpose of this study. However, these factors did not have a significant impact when users were informed about the purpose of the study, but on the other hand their familiarity with mobile device security helped them to recognise the fake QR codes. This shows that when users are well informed and become aware of social engineering they may be more impulsive in their decision making towards mobile device security.

As a result, a forensic investigation guideline was suggested in accordance with common security mistakes of mobile device users. The guideline suggested to focus on possible remnants of user activities resulted of a user impulsivity, lack of knowledge and understanding during forensics investigation. This research only briefly touched the aspect of application permissions and malware on mobile devices, therefore the future research could further investigate users' perception of application permissions on their device and the limits of intrusion that users allow for the applications and suggest an investigation guideline based on users' awareness and their ability to recognise malware on their mobile devices.

## 8. References

[1] F. N. Dezfouli, A. Dehghantanha, R. Mahmod, N. F. B. M. Sani, and S. bin Shamsuddin, "A Data-centric Model for Smartphone Security," *Int. J. Adv. Comput. Technol.*, vol. 5, no. 9, 2013.

[2] F. N. Dezfouli, A. Dehghantanha, B. Eterovic-Soric, and K.-K. R. Choo, "Investigating Social Networking applications on smartphones detecting Facebook, Twitter, LinkedIn and Google+ artefacts on Android and iOS platforms," *Aust. J. Forensic Sci.*, 2015.

[3] M. Damshenas, A. Dehghantanha, K.-K. R. Choo, and R. Mahmud, "M0Droid: An Android Behavioral-Based Malware Detection Model," *J. Inf. Priv. Secur.*, vol. 11, no. 3, Sep. 2015.

[4] H. Shewale, S. Patil, V. Deshmukh, and P. Singh, "Analysis of Android Vulnerabilities and Modern Exploitation Techniques," *Ictactjournals.in*, vol. 6948, no. March, pp. 863–867, 2014.

[5] X. Hei, X. Du, and S. Lin, "Two vulnerabilities in Android OS kernel," *IEEE Int. Conf. Commun.*, pp. 6123–6127, 2013.

[6] S. Mohtasebi, "Smartphone Forensics: A Case Study with Nokia E5-00 Mobile Phone," *Int. J. …*, vol. 1, no. 3, pp. 651–655, 2011.

[7] I B M Security and M. Pistoia, "Dynamic Detection of Inter-application Communication Vulnerabilities in Android Categories and Subject Descriptors," 2011.

[8] K. Shah, T. Shenvi, K. Desai, R. Asrani, and V. Jain, "Phishing: An Evolving Threat," vol. 3, no. January, pp. 216–222, 2015.

[9] A. Smith, "U.S. Smartphone Use in 2015 | Pew Research Center," 2015. [Online]. Available: http://www.pewinternet.org/2015/04/01/us-smartphone-use-in-2015/. [Accessed: 17-Dec-2015].

[10] R. Costin and D. Emm, "Securelist | Kaspersky Security Bulletin 2013. Malware Evolution - Securelist," *Securelist*, 2013. [Online]. Available: https://securelist.com/analysis/kaspersky-security-bulletin/57879/kaspersky-security-

bulletin-2013-malware-evolution/. [Accessed: 09-Dec-2015].

[11] M. Damshenas, A. Dehghantanha, and R. Mahmoud, "A survey on digital forensics trends," *Int. J. Cyber-Security Digit. Forensics*, vol. 3, no. 4, pp. 209–235, Oct. 2014.

[12] P. N. Ninawe and S. B. Ardhapurkar, "Design and implementation of cloud based mobile forensic tool," in *2015 International Conference on Innovations in Information, Embedded and Communication Systems (ICIIECS)*, 2015, pp. 1–4.

[13] E. R. Mumba and H. S. Venter, "Mobile forensics using the harmonised digital forensic investigation process," in *2014 Information Security for South Africa*, 2014, pp. 1–10.

[14] S. Parvez, A. Dehghantanha, and H. G. Broujerdi, "Framework of digital forensics for the Samsung Star Series phone," in *2011 3rd International Conference on Electronics Computer Technology*, 2011, vol. 2, pp. 264–267.

[15] D. Quick and K.-K. R. Choo, "Dropbox analysis: Data remnants on user machines," *Digit. Investig.*, vol. 10, no. 1, pp. 3–18, Jun. 2013.

[16] T. Y. Yang, A. Dehghantanha, K.-K. R. Choo, and Z. Muda, "Windows Instant Messaging App Forensics: Facebook and Skype as Case Studies.," *PLoS One*, vol. 11, no. 3, p. e0150300, Jan. 2016.

[17] F. Daryabar and A. Dehghantanha, "Investigation of Malware Defence and Detection Techniques," *Int. J. Digit. Inf. Wirel. Commun.*, vol. 1, no. 3, pp. 645–650, 2011.

[18] M. Damshenas, A. Dehghantanha, and R. Mahmoud, "A Survey on Malware propagation, analysis and detection," *Int. J. Cyber-Security Digit. Forensics*, vol. 2, no. 4, pp. 10–29, 2013.

[19] K. Shaerpour, A. Dehghantanha, and R. Mahmod, "Trends in Android Malware Detection.," *J. Digit. Forensics, Secur. Law*, vol. 8, no. 3, pp. 21–40, 2013.

[20] F. F. N. Dezfoli, A. Dehghantanha, R. Mahmoud, N. F. B. M. Sani, and F. Daryabar, "Digital Forensic Trends and Future," *Int. J. Cyber-Security Digit. Forensics*, vol. 2, no. 2, pp. 48–76, 2013.

[21] A. Kharraz, E. Kirda, W. Robertson, D. Balzarotti, and A. Francillon, "Optical Delusions: A Study of Malicious QR Codes in the Wild," no. December, 2012.

[22] K. Krombholz, P. Frühwirt, P. Kieseberg, I. Kapsalis, and E. Weippl, "QR Code Security: A Survey of Attacks and Challenges for Usable Security," *Hum. Asp. Inf. Secur. Privacy, Trust*, vol. 8533, pp. 79–90, 2014.

[23] A. Dehghantanha and R. Ramli, "A User-Centered Context-sensitive Privacy Model in Pervasive Systems," *IEEE*, no. 2010 Second International Conference on Communication Software and Networks, pp. 78–82, 2010.

[24] A. Aminnezhad, A. Dehghantanha, and M. T. Abdullah, "A survey on privacy issues in digital forensics," *Int. J. Cyber-Security Digit. Forensics*, vol. 1, no. 4, pp. 311–323, 2012.

[25] M. N. Yusoff, R. Mahmod, M. T. Abdullah, and A. Dehghantanha, "Mobile forensic data acquisition in Firefox OS," in *2014 Third International Conference on Cyber Security, Cyber Warfare and Digital Forensic (CyberSec)*, 2014, pp. 27–31.

[26] K. Qian, C.-T. Dan Lo, M. Guo, P. Bhattacharya, and L. Yang, "Mobile security labware with smart devices for cybersecurity education," in *IEEE 2nd Integrated STEM Education Conference*, 2012, pp. 1–3.

[27] M. Pattinson, C. Jerram, K. Parsons, A. McCormac, and M. Butavicius, "Why do some people manage phishing emails better than others?," *Inf. Manag. Comput. Secur.*, vol. 20, no. 1, pp. 18–28, 2012.

[28] J. Imgraben, A. Engelbrecht, K. R. Choo, J. Imgraben, A. Engelbrecht, and K. R. Choo, "Always connected, but are smart mobile users getting more security savvy? A survey of smart mobile device users," vol. 3001, no. December 2015, 2014.

[29] S. Mansfield-Devine, "Mobile security: it's all about behaviour," *Netw. Secur.*, vol. 2014, no. 11, pp. 16–20, 2014.

[30] Q. Do, B. Martini, and K.-K. R. Choo, "A Forensically Sound Adversary Model for Mobile Devices.," *PLoS One*, vol. 10, no. 9, p. e0138449, Jan. 2015.

[31] A. Sabeeh, "Users ' Perceptions on Mobile Devices Security Awareness in Malaysia," no. JANUARY 2011, 2015.

[32] R. Coustick-Deal, "Open Rights Group - Responding to 'Nothing to hide, Nothing to fear,'" 2015. [Online]. Available: https://www.openrightsgroup.org/blog/2015/responding-to-nothing-to-hide-nothing-to-fear. [Accessed: 18-Dec-2015].

[33] MHG, "QR Code Usage Survey: The Impact of QR Codes on Advertising Recall | MGH | Baltimore, MD- Washington, D.C.," Baltimore, MD- Washington, D.C., 2011.

[34] M. Shariati, A. Dehghantanha, B. Martini, and K.-K. R. Choo, "Chapter 19 - Ubuntu One investigation: Detecting evidences on client machines," in *The Cloud Security Ecosystem*, R. K.-K. R. Choo, Ed. Boston: Syngress, 2015, pp. 429–446.

[35] M. Shariati, A. Dehghantanha, and K.-K. R. Choo, "SugarSync forensic analysis," *Aust. J. Forensic Sci.*, vol. 48, no. 1, pp. 95–117, 2016.

[36] F. Daryabar, A. Dehghantanha, B. Eterovic-Soric, and K.-K. R. Choo, "Forensic investigation of OneDrive, Box, GoogleDrive and Dropbox applications on Android and iOS devices," *Aust. J. Forensic Sci.*, pp. 1–28, Mar. 2016.

[37] J. W. H. Bullee, L. Montoya, W. Pieters, M. Junger, and P. H. Hartel, "The persuasion and security awareness experiment: reducing the success of social engineering attacks," *J. Exp. Criminol.*, vol. 11, no. 1, pp. 97–115, 2015.

[38] M. Iqbal, M. M. Abid, M. Ahmad, and F. Khurshid, "Study on the Effectiveness of Spam Detection Technologies," *Int. J. Inf. Technol. Comput. Sci.*, vol. 8, no. 1, pp. 11–21, 2016.